\documentclass[informatics,article,accept,oneauthor,pdftex,10pt,a4paper]{mdpi} 
\firstpage{1} 
\articlenumber{18}
\doinum{10.3390/informatics4030017}
\pubvolume{4}
\pubyear{2017}
\copyrightyear{2017}
\externaleditor{Academic Editors: Achim Ebert and Gunther H. Weber}
\history{Received: 27 May 2017; Accepted: 24 June 2017; Published: 27 June 2017}

\pdfoutput=1

\pdfpageattr {/Group << /S /Transparency /I true /CS /DeviceRGB>>}

\Title{TOPCAT: Desktop Exploration of Tabular Data for Astronomy and Beyond}

\Author{Mark Taylor$^{}$}

\AuthorNames{Mark Taylor}

\address[1]{%
             H.~H.~Wills Physics Laboratory,
             University of Bristol, Tyndall Avenue, Bristol,
             BS8 1TL, UK; m.b.taylor@bristol.ac.uk}

\corres{Correspondence: m.b.taylor@bristol.ac.uk}

\abstract{TOPCAT, the Tool for OPerations on Catalogues And Tables,
is an interactive desktop application for retrieval, analysis and
manipulation of tabular data, offering a powerful and flexible
range of interactive visualization options amongst other features.
Its visualization capabilities focus on enabling interactive exploration
of large static local tables---millions of rows and hundreds of columns can easily be handled
on a standard desktop or laptop machine, and various options
are provided for meaningful graphical representation
of such large datasets.
TOPCAT has been developed in the context of astronomy,
but many of its features are equally applicable to other domains.
The~software, which is free and open source, is written in Java,
and the underlying high-performance visualisation library
is suitable for re-use in other applications.
}


\begin{document}

\section{Introduction}
\vspace{-6pt}

\subsection{Source Catalogues}

Astronomy is a discipline with a long history of
collecting, storing and analysing data.
This~comes in various forms, including images, spectra and time
series, but one of the most important is the {\em source catalogue},
a list of observed astronomical objects such as stars or galaxies,
each with a fixed set of features.
These features are mostly numeric and typically include quantities
such as central sky coordinates, brightness in one or several wavebands,
apparent size, ellipticity parameters and so on.

A catalogue is thus
naturally represented as a table with a certain number of
rows (one for each observed object) and columns (one for each feature).
An early example is the Catalog of Nebulae and Star Clusters
published by Charles Messier in 1781 \cite{1781cote.rept..227M},
containing data in six columns for 103~celestial objects,
each observed individually by eye.
More recent examples are often, though not always, considerably larger;
the features for modern catalogues are extracted automatically from
image data obtained by highly sophisticated telescope instrumentation,
and for the largest sky surveys can run to hundreds of columns
such as the Sloan Digital Sky Survey
\citep{2002AJ....123..485S},
and/or the order of a billion rows
such as the Gaia mission
\citep{2016A&A...595A...2G}.
These numbers, of course, are expected to rise in the future,
for instance in upcoming experiments such as
ESA's Euclid satellite (\url{http://sci.esa.int/euclid/})
and the 
Large Synoptic Survey Telescope (\url{https://lsst.org/}).
Many other catalogues however may contain only a few tens or
thousands of objects identified as a particular astronomical type
or of interest for a particular study.
In some cases catalogues can be represented by a single table,
in others as complex relational~databases.

A number of technologies exist for accessing such data,
including bulk download of full or partial tables in
domain-specific
(VOTable \citep{2013ivoa.spec.0920O}, FITS \citep{2001A&A...376..359H})
or generic (Comma-Separated Value) file formats,
and remote access to relational databases using SQL-like query languages.
A family of ``Virtual Observatory'' standards
(see e.g.,\ \citep{2010ivoa.rept.1123A,2010ivoa.spec.0327D,2008ivoa.specQ0222P}),
developed since 2002 by a group of interested parties known as the 
International Virtual Observatory Alliance
(IVOA) (\url{http://www.ivoa.net/}),
enables standardised access to many thousands of such catalogues,
which are mostly available without usage restrictions,
hosted by a large network of data servers around the world.

\subsection{TOPCAT Application}

TOPCAT (\url{http://www.starlink.ac.uk/topcat/}),
the Tool for OPerations on Catalogues And Tables~\citep{2005ASPC..347...29T},
is a desktop Java GUI application for retrieval, analysis,
and manipulation of tables.
It has been developed in the context of astronomy
and the Virtual Observatory, with the
aim of providing a toolkit for astronomers to perform
all the mechanical operations they routinely require on
catalogues, so they can focus on extracting scientific
meaning from this hard-won data.

It has been under more or less continuous development since 2003,
and is in 2017 a mature application
with an active user base in the thousands,
spread over six continents,
including undergraduates, amateur astronomers, and research scientists.
A number of factors have contributed to its popularity in the 
astronomy community,
including astronomy-specific capabilities such as celestial
coordinate system handling, table joins using sky positions,
and Virtual Observatory data access,
as~well as more generic items such as its
powerful expression language and support for large datasets,
alongside a responsive development model,
high-quality support,
ease of installation
and a relatively shallow initial learning curve.

This paper however describes just one aspect of TOPCAT's operation,
its capabilities for exploratory visualisation,
especially as applicable to generic
(not necessarily astronomical) tabular data.
Its distinctive combination of features
compared to other visualisation applications
includes:
\begin{itemize}[leftmargin=*,labelsep=5.5mm]
\item the ability to work with large datasets,
      without any special preparation of the data
      or prior assumptions about the visualisations required
\item provision of many options to explore high-dimensional data,
      that can be adjusted interactively with rapid visual feedback
\item meaningful representation of both high and low density
      regions of very large point clouds
\end{itemize}

The rest of the paper discusses these capabilities and
outlines some of the underlying implementation.
Section~\ref{sec:overview} describes TOPCAT's relatively unsophisticated
approach to data access which can nevertheless,
by using robust technologies such as file mapping, deliver high
performance results, as well as providing a platform that is
easy to deploy and install.
Sections \ref{sec:pointcloud} and \ref{sec:otherplot}
explore the visualisation capabilities offered,
principally representation of point clouds in one, two and three dimensions,
including the optionally weighted ``hybrid density map/scatter plot''
which provides a unified view of high and low density regions
in very crowded plots;
the use of {\em linked views} for exploring high-dimensional data
is also discussed.
Sections \ref{sec:configui} and \ref{sec:otherui}
examine the difficult issue of providing a comprehensible
user interface to control the highly configurable plots on offer,
including mention of the command-line interface STILTS.
Section~\ref{sec:performance} goes into some detail about the
implementation of certain performance-critical parts of the code
and
Section~\ref{sec:nonastro} gives a few examples of its,
currently minority,
use in fields other than astronomy.
Sections \ref{sec:availability} and \ref{sec:conclusions}
conclude with information about availability of the~software.

The visualisation framework described here corresponds to that introduced
in version 4 of the application, released in 2013.

\section{Application Overview}
\label{sec:overview}
\vspace{-6pt}

\subsection{Data Access}
\label{sec:dataaccess}

TOPCAT uses a traditional model of data access for visualisation,
in which the user identifies and retrieves
to local storage one or more static tables, and then works with them.
This makes it unsuitable for direct visualisation of extremely large
datasets, but it turns out in most cases to be possible
for users working with very large astronomical catalogues
to preselect and download a subset of interest,
by restricting for instance to a given sky region or class of
astronomical object; TOPCAT provides a wide range of astronomy-specific
options for selective acquisition of such data.
In many other cases, users will be concerned with much smaller
catalogues where data volume is not an issue.

Given this approach, it is important to be able to access large
data files on local disk efficiently.
TOPCAT's preferred input format is the FITS binary table
\citep{2001A&A...376..359H}.
This binary format lays out columns and rows in a predictable
pattern on disk, so that files can be {\em mapped\/} into memory
for sequential or random access,
giving effectively instant load time and without encroaching on
Java's limited heap space; for more explanation of this technique and its
benefits in this context see \cite{2008ASPC..394..422T}.
However, other formats such as the more common Comma-Separated Values (CSV)
are also supported.
In this case direct file mapping is not useful,
but for large CSV tables the data is copied on load into a temporary
binary file which can itself be mapped,
allowing similar access but with a significant load time.
Another option is to use TOPCAT to convert from CSV to FITS format
before use.

We note that more sophisticated data access models are in use by other
visualisation applications, for instance
running the computation on a remote data-hosting server
and transmitting only the resulting images to be displayed
in a browser or other desktop application 
(e.g.,\ the Gaia archive visualisation service \cite{gaiaviz}),
or retrieving and caching relevant data subsets
on demand for local rendering as required
by user navigation actions
(e.g.,\ Aladin-Lite \citep{2014ASPC..485..277B})
or moving the user to the data using centralised
high-performance visualisation facilities which may include
large-scale display hardware alongside High Performance Computing capability
(e.g.,\ \cite{2017ApJS..228...19C}).
Such techniques can avoid wholesale transfer of an
impractically large dataset without requiring the user to
identify any particular subset of interest,
and can deliver excellent interactive experiences.
They also however suffer from some limitations.
Visualisations which are intrinsically data intensive will
have large resource requirements, consuming centralised
resources while in progress which may be expensive
and scale poorly to large numbers of users.
In some cases, efficient server-side visualisation makes use of
pre-computed data structures such as indexes
or hierarchical multi-resolution maps,
which may be expensive to compute but
once in place can support rapid navigation and interaction.
This can work well, but it generally requires prior information
(or assumptions) about what visualisations are going to be required.
In the case that there are many columns, and a user may want to
plot not just any pair of columns against each other, but
arbitrary functions based on available columns,
it is typically not possible to ensure that
the appropriate pre-calculated data structures are in place.
As a general rule, coordinating client and server software adds a layer of
complexity which can make software development slower and harder,
and often impacts reliability.
These techniques are also not suitable in the
absence of network connectivity.
TOPCAT's low-tech approach on the other hand has the benefits
of reliability, network independence and above all
flexibility in terms of the visualisation options available.

\subsection{Usage Model}

As well as its traditional approach to data access,
TOPCAT supports a straightforward usage model:
it is a standalone application running CPU-based code,
and suitable for use on low-end desktop or laptop computers.
The visualisation is multi-threaded to maintain GUI responsiveness,
but does not currently distribute the bulk of its computation
across multiple cores for efficiency (though it may do so in future).
If GPUs are present they are not used except for normal graphical
operations.

This generally low-tech approach can nevertheless deliver
performant interactive visualisation for quite large datasets,
and has the benefit that barriers to use are low.

\subsection{Expression Language}

One of the features of TOPCAT not directly related to visualisation
is its provision of a powerful expression 
language
which allows evaluation of simple or complex expressions involving
column names.
In general, wherever a coordinate is supplied for plotting,
either a column name or an expression can be used, making it
very easy to plot arbitrary functions or combinations of columns.
The expression language can also be used to define row selections~algebraically.

The implementation of this feature is based on
JEL, the Java Expressions Language,
available from \url{https://www.gnu.org/software/jel/}.

\section{Visualising Point Clouds}
\label{sec:pointcloud}

A source catalogue may contain tens or hundreds of columns,
and interesting relationships may be lurking between pairs or
higher-order tuples of these features.
Often, an interesting result is to identify a subset
of rows occupying a particular region
in some multidimensional parameter space whose axes may
be table columns, or linear or non-linear combinations of columns.
Physically, this~corresponds to identifying a sub-population of
observed astronomical objects sharing some physical characteristics,
for instance a group of stars formed from the same primordial dust cloud.
TOPCAT does not attempt to provide automated support for
discovering such relationships, for instance by implementing
data mining algorithms.
Instead, it aims to provide the user with a flexible toolkit of
options to display different aspects of the data,
in order to pick out trends, associations or interesting outliers
by eye.

Many, though not all, of these options are variations on the
theme of a scatter plot in two or three dimensions
of some point cloud in multi-dimensional space.
A scatter plot can be an excellent tool for presenting a
relationship between known variables, but it presents two main problems.
First, if the dimensionality of an association is greater
than that of the plot, the association may be masked.
Second, if the number of points is large compared to the area
on which they are plotted, data can be obscured.
These issues can to some extent be addressed by providing a range
of plotting and interaction options,
and are discussed in the following subsections.

\subsection{High-Dimensional Plots}
\label{sec:highdim}

To represent a relationship between two variables by plotting
points on a two-dimensional plotting surface is straightforward.
This can be extended to three dimensions by using various techniques
for representing points in a 3-d space, though visual interpretation
tends to be harder in this case.
TOPCAT supports both options, though the 3-d representation is
at present restricted to a 2-d projection whose 3-d nature only
becomes apparent from user interaction with the mouse
(rotation, zooming,~navigation).

To visualise a higher-dimensional relationship however,
spatial positioning is not enough,
so~TOPCAT provides various ways to modify the representation
of each point according to additional features.
Distinct sub-populations can be identified using markers of different
colours, sizes or shapes, individual points can be labelled
with per-object text labels, and additional numeric features
can be encoded using:
\begin{itemize}[leftmargin=*,labelsep=5.5mm]
\item colour from a selected colour map
\item marker size
\item X/Y marker extent
\item error bars aligned with the axes
\item vector with magnitude and orientation
\item ellipse primary/secondary radius and orientation
\end{itemize}

\begin{figure}[H]
\begin{center}
\includegraphics[width=0.50\textwidth]{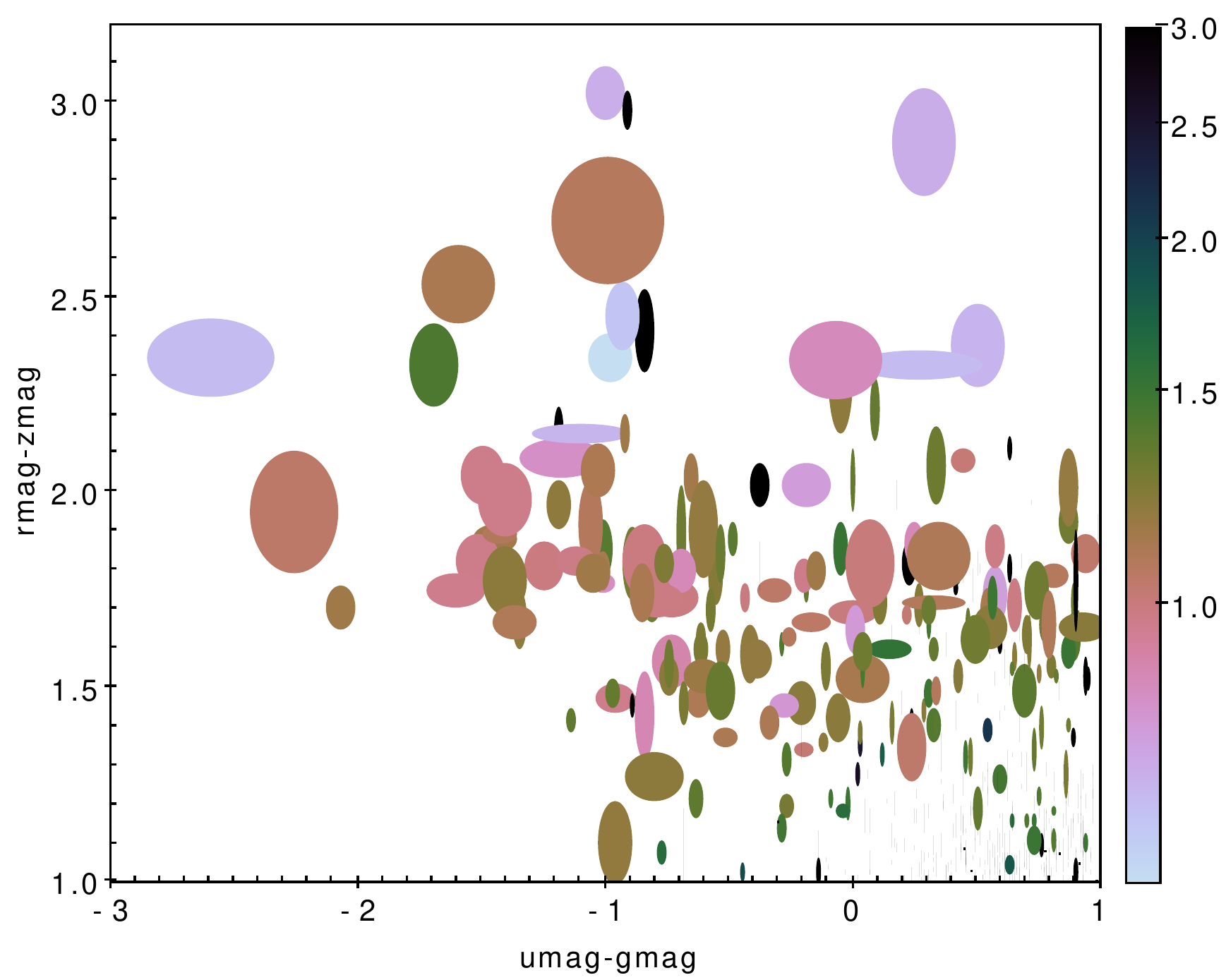}
\includegraphics[width=0.40\textwidth]{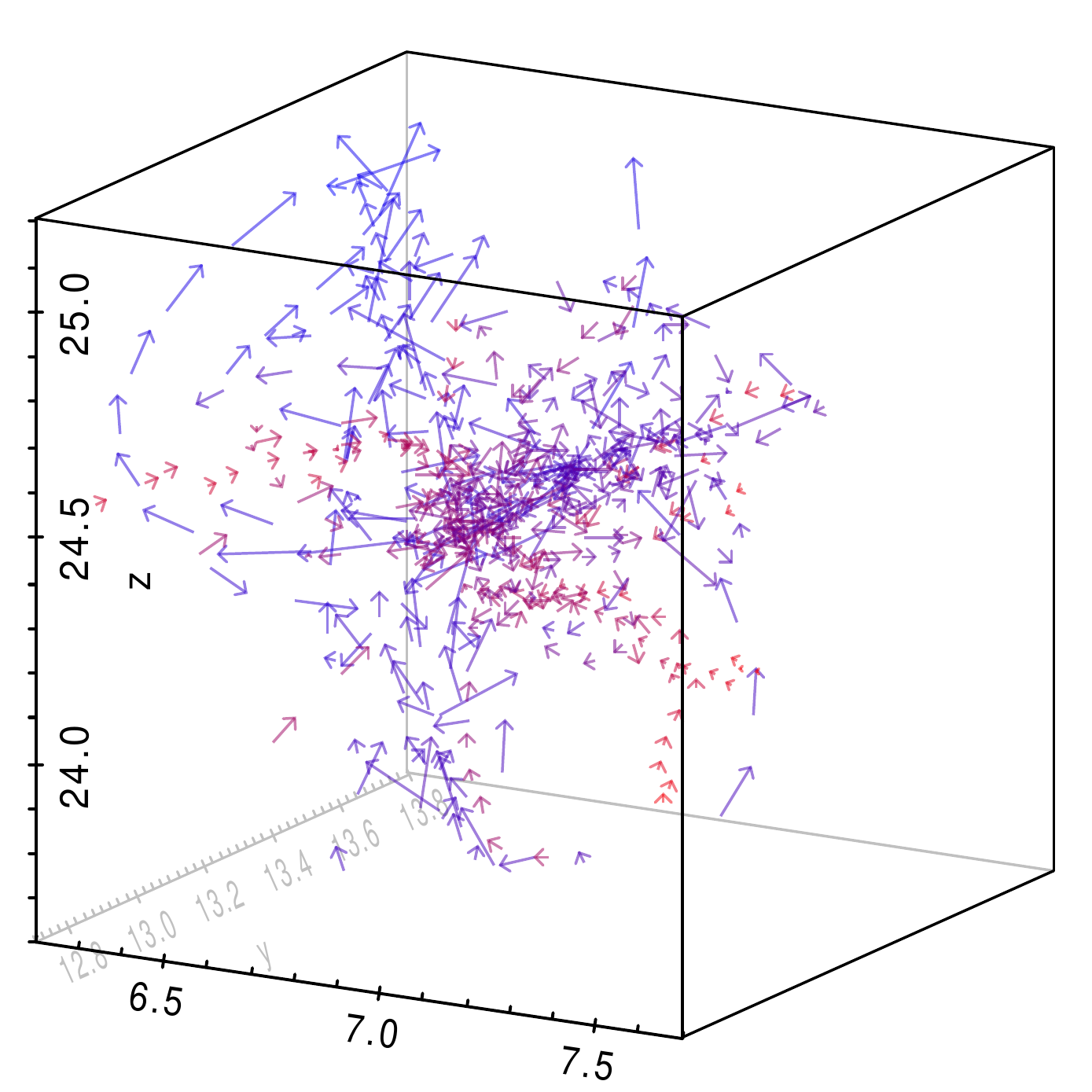}
\end{center}
\caption{
  Options for high-dimensional visualisation.
% The left hand figure uses marker colour and size to indicate additional
% features, as well as identifying two different subsets by marker shape.
  The left hand figure uses marker colour and shape to indicate
  three non-positional numeric features.
  The right hand figure
  uses arrows to represent points in six-dimensional phase space,
  the positions and velocities of simulated galaxies
  \citep{2005Natur.435..629S};
  redshift is additionally shown by colour-coding.
}
\label{fig:highdim}
\end{figure}

The user can combine these options freely;
some examples are shown in in Figure~\ref{fig:highdim}.

In principle quite a large number of features can be encoded
in this way,
for instance one could represent seven dimensions on a 2-d
scatter plot by marking coloured ellipses with text labels.
In~practice however, especially if the number of points is large,
there are limits to what is visually~comprehensible.

\subsection{Subset Selection and Linked Views}
\label{sec:linked}

An alternative approach to understanding high-dimensional data
is to extract sub-populations for further examination.
A common workflow is to make a scatter plot in one parameter space,
identify by eye a subset of points falling within a sub-region
of that space and then replot the subset in a second parameter space
to reveal some relationship evident in the subset but not the
full dataset.  This~sequence may often be iterated to narrow down
a sub-population of interest.
This can be seen as a somewhat crude way to identify clusters
that would be evident in a much higher dimensional space by
combining multiple 2-dimensional views,
but it is a powerful technique, and falls into the category
of {\em linked views}
\citep{Tukey77,2012AN....333..505G}.

\begin{figure}[H]
\centering
\raisebox{0.4\textwidth}{ (\textbf{a})}
\includegraphics[width=0.4\textwidth]{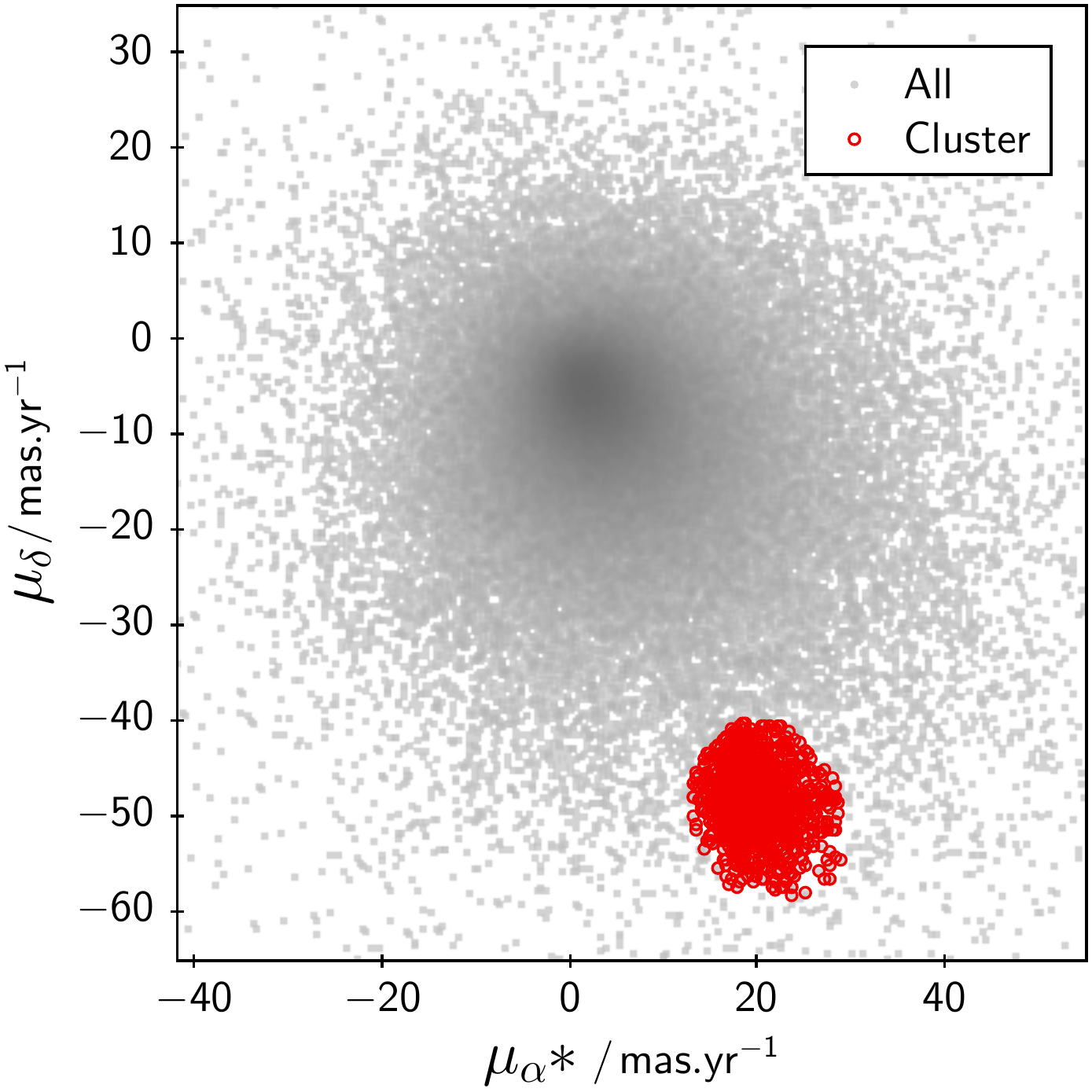}
\hspace*{0.0\textwidth}
\raisebox{0.4\textwidth}{ (\textbf{b})}
\includegraphics[width=0.4\textwidth]{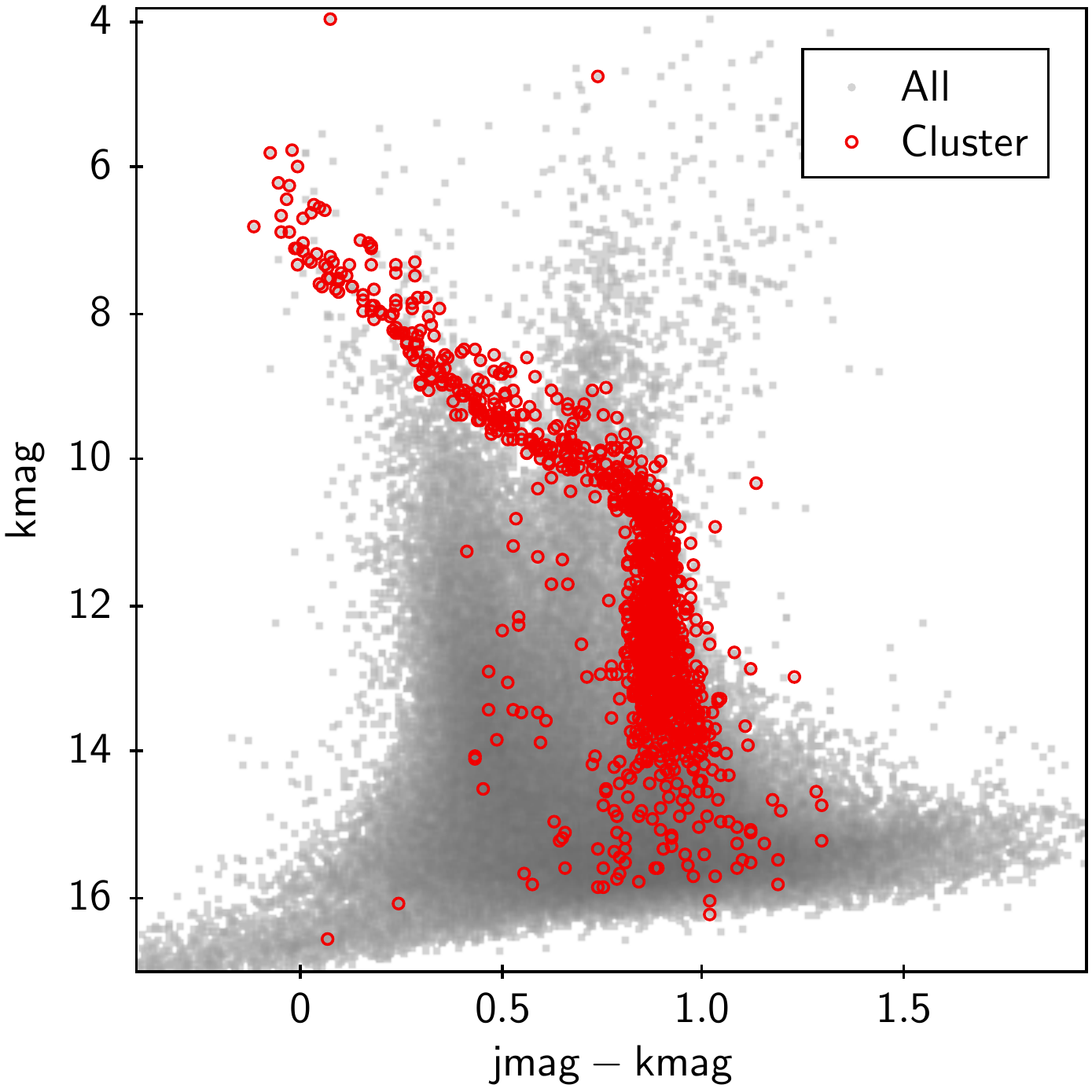}
\caption{
 Linked views for row subsets.
  The user has identified the region in plot  (\textbf{a})\/ by dragging
  the mouse, and the same subset of rows shows up in an interesting
  subregion of the different plot in  (\textbf{b}).
  In~this case each point represents a star in the region of the Pleiades
  open cluster \citep{2017A&A...600L...4A};
  (\textbf{a})\/ shows apparent velocity across the sky,
  while  (\textbf{b})\/ characterises stellar classification.
  Those stars with similar motion, identified as the ``Cluster'' subset
  in  (\textbf{a})\/, were formed in the same environment and therefore
  have similar physical characteristics, hence
  trace out a distinct path in  (\textbf{b}).
}
\label{fig:pleiades}
\end{figure}

TOPCAT allows the user to define subsets in various graphical
and non-graphical ways, one of the most powerful being to
drag out with the mouse an arbitrary shape over a region of a plot.
Such subsets, once defined, can be plotted separately in any other
plot of the same table, and also distinguished for other processing
operations.
An example is given in Figure~\ref{fig:pleiades}.

\subsection{Row Highlighting and Linked Views}
\label{sec:activate}

Another aspect of linked plots in TOPCAT is that when the user
highlights a plotted point by clicking on it,
any point in other visible plots
representing the same row is automatically highlighted.
At the same time, if the underlying data is displayed in an
table browser window, the corresponding row is highlighted so that the
data values in all columns can easily be seen.  The same operation
works in reverse, so clicking a row in the table browser window will
highlight any corresponding points in currently visible plots.

The application can also be configured so that some
{\em Activation Action\/} takes place when a row is selected by
user action in either of these ways.
A typical activation action might be to display an image
associated with the table row in question,
for instance the original photograph that supplied the data,
e.g.,\ available from a URL column in the table.
TOPCAT can perform basic image display internally,
or~communicate with external specialised display
applications in order to achieve this kind of thing.
This coordinated row-highlighting behaviour can be especially
useful for investigating outliers.

\begin{figure}[H]
\begin{center}
\includegraphics[width=0.8\textwidth]{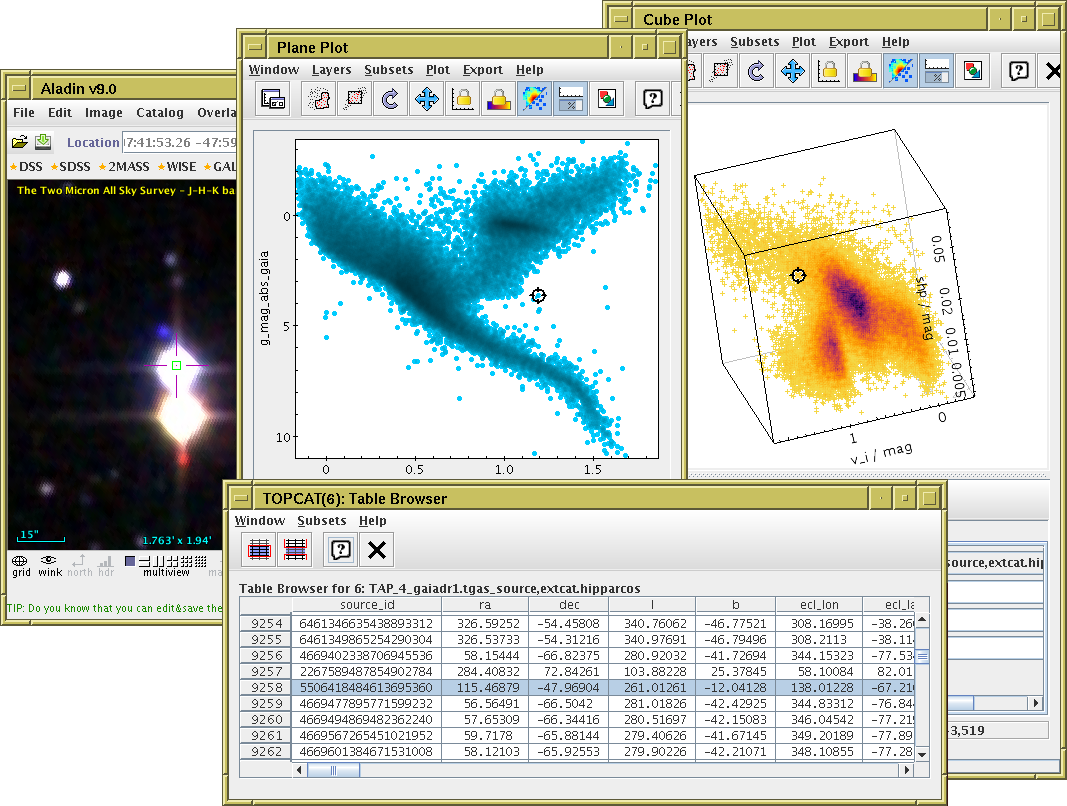}
\end{center}
\caption{
  Linked highlighting of table rows.
  A user has clicked on an outlier in the 2-d plot (\textbf{center}),
  highlighting it with a ``target'' cursor.
  This automatically causes the same row to be highlighted in other ways:
  the target cursor marks the relevant point on the 3-d plot (\textbf{right})
  and the row is flagged in the table data browser (\textbf{bottom}).
  In this case TOPCAT has also been configured to communicate with the
  external image display application Aladin
  \citep{2000A&AS..143...33B} (\textbf{left}),
  which is caused to display a sky image
  corresponding to the highlighted row.  
  This interaction makes it easy to see here that the relevant star
  is very close to another one; it is likely that this
  contamination of sources has led to a spurious brightness value,
  resulting in its anomalous position in the 2-d plot. Data show a Gaia-Hipparcos colour magnitude diagram and 2MASS colour image.
  % SELECT
  %    gaia.*,
  %    gaia.phot_g_mean_mag+5*log10(gaia.parallax)-10 AS g_mag_abs_gaia,
  %    gaia.phot_g_mean_mag+5*log10(hip.plx)-10 AS g_mag_abs_hip,
  %    hip.*
  % FROM gaiadr1.tgas_source AS gaia
  % INNER JOIN extcat.hipparcos AS hip ON gaia.hip = hip.hip
  % WHERE gaia.parallax/gaia.parallax_error >= 5
  %   AND hip.plx/hip.e_plx >= 5
  %   AND hip.e_b_v > 0.0
  %   AND hip.e_b_v <= 0.05
  %   AND 2.5/log(10)*gaia.phot_g_mean_flux_error/gaia.phot_g_mean_flux <= 0.05;
  % 2d plot: b_v vs. g_mag_abs_gaia
  % 3d plot: g_mag_abs_hip vs. v_i vs shp
  % Aladin: 2MASS JHK bands (RGB colour)
}
\label{fig:activation}
\end{figure}

The communication with external display applications mentioned above,
if required,
can be done by invoking methods of Java classes supplied at runtime,
or invoking system commands, or using the 
messaging protocol SAMP \citep{2015A&C....11...81T}
implemented by a number of astronomy tools,
including
Aladin~\citep{2000A&AS..143...33B},
SAOImage DS9 \citep{2003ASPC..295..489J},
Astropy \citep{2013A&A...558A..33A} and others.
This latter option in fact provides for inter-process
exchange of both single rows and row subsets,
allowing linked views {\em between\/} cooperating
applications, as well as just within TOPCAT.

For an example of all this in action, see Figure~\ref{fig:activation}.

\subsection{High-Density Plots}

TOPCAT aims to be able to explore tables containing many rows.
The answer to the question, ``how many?'', is of course constrained
by available resources of computation and user time,
but in general the target is: as many as possible.
The larger the dataset that can be explored interactively,
the fewer decisions the user will need to take in pre-selecting data,
and the more relationships are potentially available for discovery.

There are two main aspects to consider when attempting to satisfy
this requirement.
The more obvious concerns resource usage: 
will the computation require more memory than is available,
and~will it be fast enough to provide a fluid experience?
These questions are discussed in Section~\ref{sec:performance}.
But there is also a question of what constitutes a visually
faithful and comprehensible representation of a very large
number of points.
In particular, how can one represent a scatter plot when
the number of points to plot exceeds the number of pixels available?

\begin{figure}[H]
\begin{center}
\includegraphics[width=0.8\textwidth]{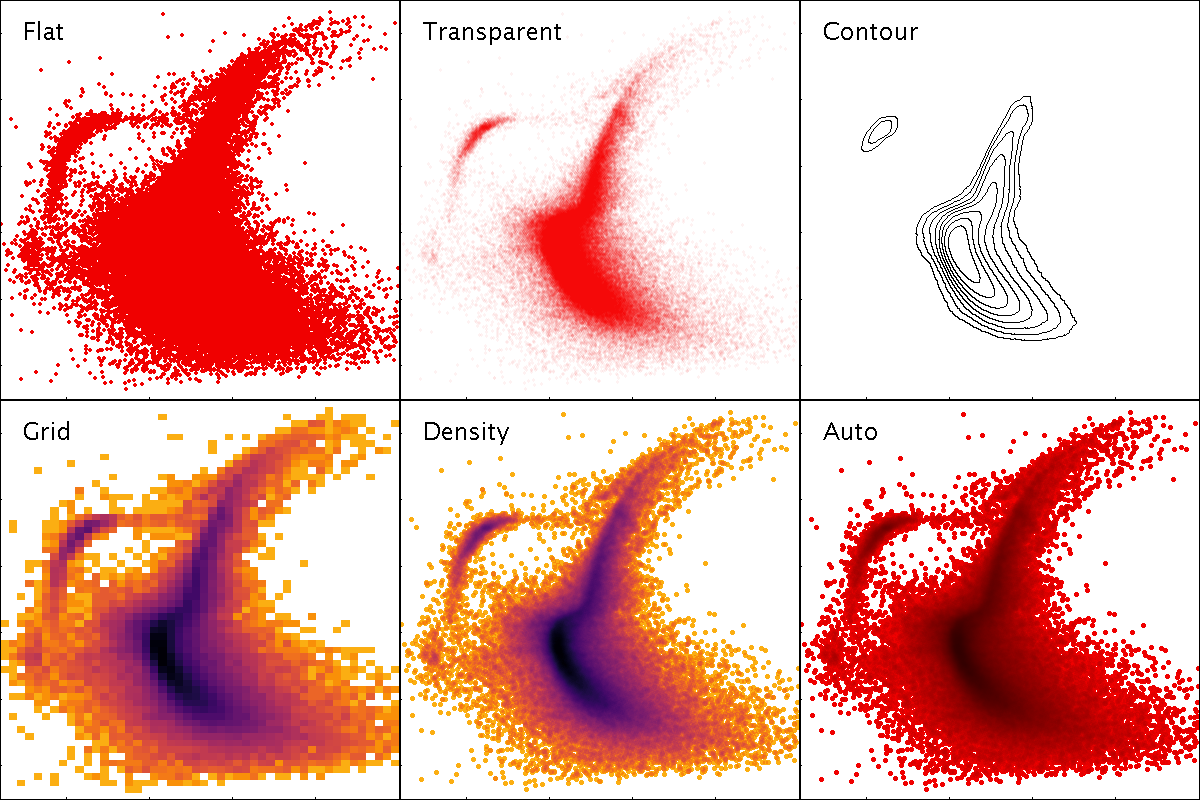}
\end{center}
\caption{
  Various representations of a 2-dimensional point cloud available
  within TOPCAT.
  {\em Flat\/} simply plots markers at each point, obscuring the density
  structure.
  {\em Transparent\/} plots partially transparent markers, giving some
  indication of overdense regions, however regions over a certain density
  threshold are saturated.
  {\em Contour\/} is a somewhat smoothed contour plot.
  {\em Grid\/} is a 2-d histogram on a grid of fixed size rectangular bins;
  various options are available for combination within bins including
  mean, median, min, max etc.
  Density structure is clear, but resolution is lost and
  outliers are poorly represented.
  {\em Density\/} is a hybrid density map/scatter plot with a configurable
  colour map, showing both high-density structure and individual outliers.
  {\em Auto\/} is a standard profile of the hybrid map used by default,
  with~a fixed colour map scaling that fades from a
  dataset's chosen colour to black;
  multiple overplotted datasets can be distinguished by using
  different base colours.
  Data represent a $V$  vs. $B$-$V$ colour-magnitude diagram of
  139,000 stars from the globular cluster $\omega$ Centauri
  \citep{2009A&A...493..959B}.
}
\label{fig:plotmodes}
\end{figure}

Simply plotting the points as opaque markers loses information
where there are many points per pixel, since it is not possible to
see how many points are overplotted.
A number of options exist to address this,
such as painting partially transparent markers,
drawing contours,
or binning data to generate colour-coded two-dimensional histograms
(also known as density maps).
Very~often for source catalogues however, the outliers are
just as important as the statistical trends, so both the
low and high density regions of the plot must be represented
faithfully.  Contour plots and density maps do not work well for
low-density regions, while transparent points are suitable
for small variations in density but lose information at
one or both ends of the density spectrum if there is a large density range.
Use of a density map or contour plot also inhibits the
row highlighting behaviour described in Section~\ref{sec:activate},
since single points are not represented.
To address this issue, a~hybrid density map/scatter plot
has been introduced,
which is a convolution of a single-pixel density map
with a shaped marker, and represents both high and low density regions
of the plot well in the same display,
resembling a smoothed density map at high density and a normal
scatter plot at low density.
This hybrid representation, which is the default plotting mode,
also works particularly well when navigating a large point cloud:
zooming in turns a high-density into a low-density region,
so the plot transitions smoothly from a density map to a normal scatter plot.
It is described in more detail in \cite{2014ASPC..485..257T}.
TOPCAT provides all these options and others for plotting large
and small point clouds in two or three dimensions.
Some examples are shown in Figure~\ref{fig:plotmodes}.

\begin{figure}[H]
\begin{center}
\includegraphics[width=0.9\textwidth]{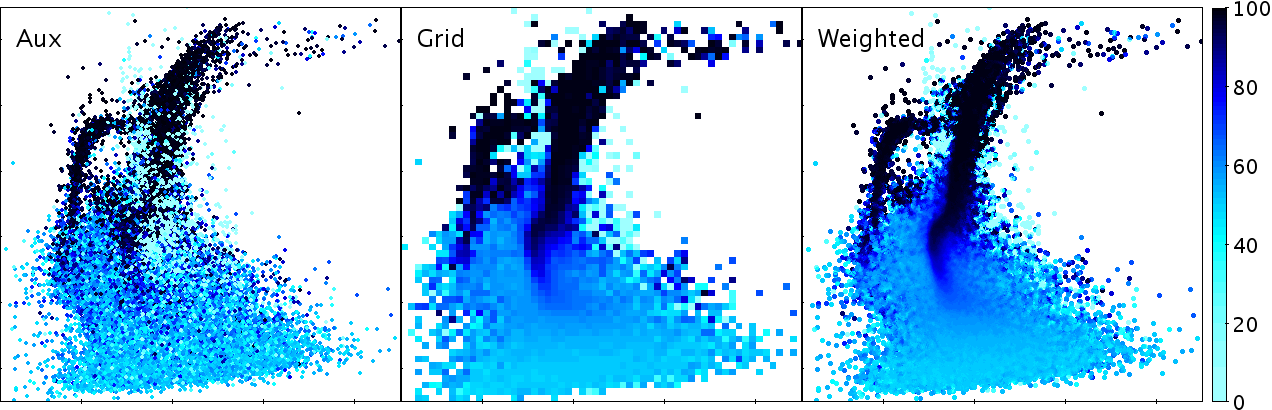}
\end{center}
\caption{
  Various representations of a 2-dimensional point cloud with a
  third dimension indicated by colour.
  {\em Aux\/} paints opaque coloured markers, which works for low-density
  regions, but at high density the appearance is noisy and
  depends on row sequence,
  since points painted later obscure earlier ones.
  {\em Grid\/} is a 2-d histogram weighted by the third coordinate;
  points in the same bin are averaged, but some spatial resolution is lost.
  {\em Weighted\/} is a hybrid density map/scatter plot but with pixel bins
  weighted by the third coordinate,
  providing clarity in both low and high density regions.
  The~weighting combination method in the latter two cases is configurable;
  in this case the median has been used.
  Data are as for Figure~\ref{fig:plotmodes} with the third coordinate
  indicating probability of cluster~membership.
}
\label{fig:weight}
\end{figure}

Colour-coding points to express extra dimensions as discussed in
Section~\ref{sec:highdim} presents additional issues in high-density
regions, since multiple colours may be overplotted in the same pixel.
To~address this problem, TOPCAT offers a weighted generalisation
of the hybrid density map/scatter plot,
as~illustrated in Figure~\ref{fig:weight}.

\subsection{Navigation}

It is impossible to understand all the information in a point
cloud consisting of millions of points from a static image,
particularly if it is presented on a grid of, say, 100,000 pixels.
But it is often possible to identify what might be a region of
interest from a plot that represents the density structure and
outliers appropriately, and zoom in on such a region for closer
inspection, down to the features of individual objects.
Interactive navigation of plots is therefore a crucial feature for
exploratory~visualisation.

The basic user interface for navigating 2-dimensional plots is fairly
straightforward,
namely that dragging a mouse around the screen will drag the plot with it,
while rolling a mouse scroll wheel will zoom in or out
around the current mouse position.
Plots in TOPCAT in many cases have no natural aspect ratio,
so various options are provided for anisotropic zooming:
dragging the middle button will drag out a ``window''
rectangle with any required aspect ratio
to become the new field of view, while
dragging the right button stretches or shrinks the field of view
in the X and Y directions independently according to the drag position.
It is also possible to control the X or Y fields of view separately
by performing the windowing or stretching gestures near the
relevant axis.
Keyboard modifiers are available for mice lacking all three buttons
or a scroll wheel.
The isotropic drag/zoom navigation gestures are generally intuitive for
users of other GUI applications.
The anisotropic zooming capabilities offer very flexible
2-d navigation, but they are less intuitive and advertising them well
is difficult, so many users may not be aware of their proper use.

Allowing navigation through a 3-d scatter plot is a more difficult
user interface problem.
In this case, TOPCAT uses the default-button drag to rotate the
visualised cube, and the mouse wheel to zoom in or out around the cube center.
These gestures are intuitive.
Translating the volume within the cube however is harder to do.
Some mouse gestures are assigned to drag and stretch along the
cube plane most nearly parallel to the screen projection plane,
but it is difficult to use these to zoom in on a region of interest.
More useful is the right-click; this takes the point under the
current cursor position and translates it to the center of the cube,
so that subsequent zoom actions will zoom in and out around it.
However, since in 3-d
the cursor position represents a line of sight rather than
a unique point, this~presents a problem: which of the
positions under the cursor should be the new plot center?
To break this degeneracy, the point chosen is the center of mass
(mean depth) of all the points plotted along the line of sight.
The effect is that when clicking on a single point, the point
position is used, while when clicking on a dense region, the
center of the region is used.
This re-centering navigation works very well in practice for
navigating to regions of interest in 3-d point clouds,
though again the UI may not be obvious to all users.

Visual feedback is given as quickly as possible in all these cases.
Typically the screen is updated at better than 10 frames per second up to
a million or so rows for one of the standard scatter plots;
for~larger datasets or more complex plots
the response may be more sluggish.
To improve user experience, an adaptive ``sketch'' mode
is in effect by default.
If refreshing a frame takes more than a certain threshold time
(0.25 s),
fast intermediate plots 
based on a subsample of the data
are drawn while a navigation action is in progress,
the subsampling fraction being chosen depending on how long the
full plot appears to take.
When the user has stopped dragging/zooming and enough time has
elapsed for the full plot to be drawn,
the display is refreshed from the full dataset.
In most cases this gives a good compromise between responsive
and accurate behaviour, though for certain plot types
the intermediate sketched frames can prove confusingly
different from the final, correct,~frame.

\section{Other Plot Types}
\label{sec:otherplot}

\begin{figure}[H]
\centering
\includegraphics[width=0.8\textwidth]{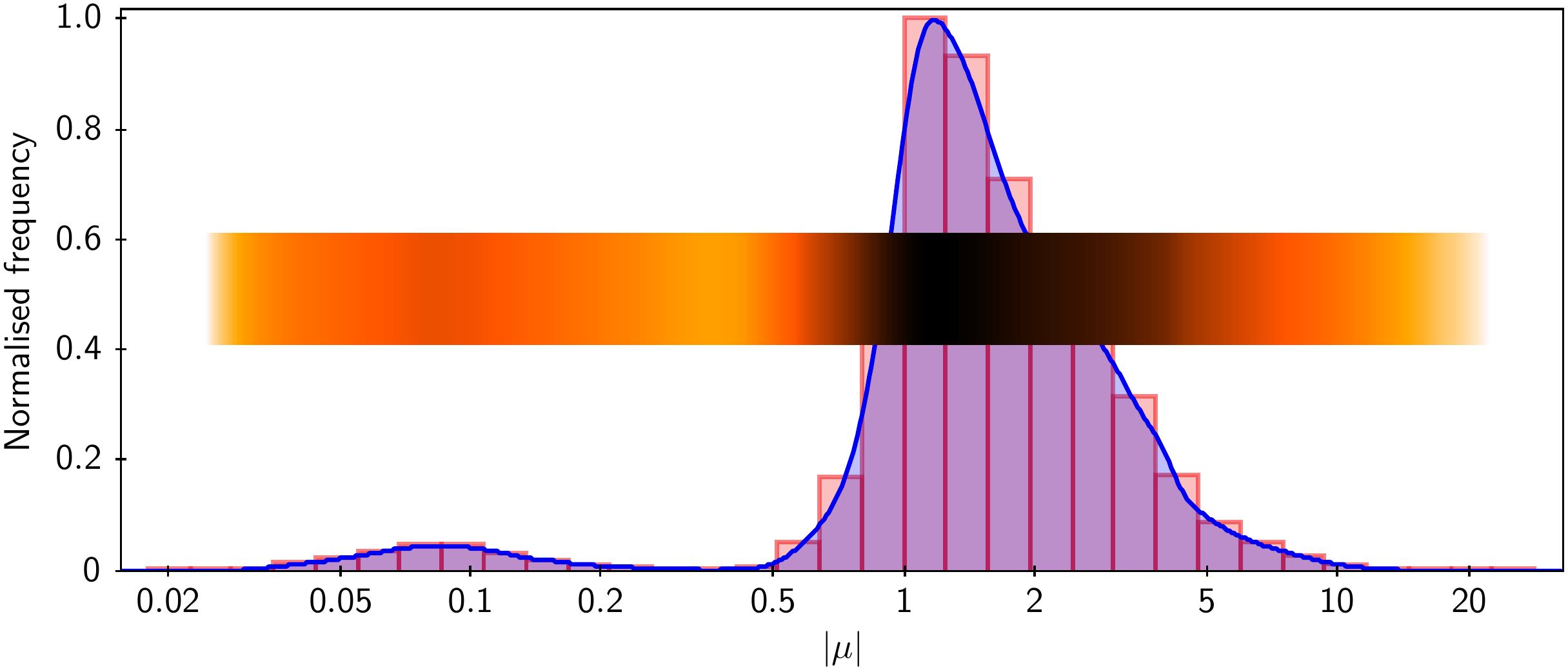}

\caption{Some histogram-like plots.
  Shown here are three different representations of the same
  one-dimensional dataset:
  a traditional histogram,
  a Kernel Density Estimate which uses a smoothing kernel
  to avoid the quantisation implied by histogram binning,
  and a ``densogram'' that represents point density
  using a colour bar.
}
\label{fig:histo}
\end{figure}

Although the main focus of the visualisation in TOPCAT is representing
point clouds of various sorts, it has other visualisation capabilities too.
One important category is depicting weighted or unweighted frequency data,
which is a kind of point cloud in one dimension.
The most common visualisation for this kind of data is a histogram,
but a number of variations are available including
smoothed representations with choices of fixed-width or adaptive
smoothing kernels with various functional forms,
a range of normalisation types, cumulative binning,
control over bin width and phase, etc.
Some of the options are illustrated in Figure~\ref{fig:histo}.

A number of other more specialised plot types are also offered,
for instance analytic function plotting,
line plots to trace samples against an independent variable,
axes annotated with time coordinates,
spectrograms,
and some fairly basic data fitting algorithms.

There are also comprehensive facilities
for plotting data with positions specified by latitude and longitude
on the celestial sphere, including
a range of sky projections,
sky coordinate system grid annotations,
and binning schemes suitable for spherical geometry.
These sky plotting capabilities may also be used for
data situated on other spheres,
such as (an approximation to) the surface of the Earth.

\section{Configuration User Interface}
\label{sec:configui}

TOPCAT currently provides around thirty different plot layer types
(fixed shape marker, variable size marker, error ellipse, histogram, ...)
with seven different shading modes
(flat, auto, transparent, weighted, ...),
on half a dozen different plot geometries
(2-d and 3-d Cartesian, spherical polar, celestial sphere, ...).
Each of these options has typically 5--10 associated configuration variables.
All of these options have been introduced to support
anticipated, and in many cases actual, requirements for
making sense of real user data.

This offers a great deal of flexibility for specifying a visualisation,
but equally presents a serious problem of complexity:
how does the user, especially the non-expert user,
navigate all these options to look at some data?
Packaging this flexibility in a comprehensible and usable GUI
is perhaps the single most difficult problem
in developing an application of this kind.

Addressing this from the user perspective, the application
is designed on the principle that the user should always see
some kind of reasonable plot with minimal effort.

\begin{figure}[H]
\begin{center}
\includegraphics[width=0.416\textwidth]{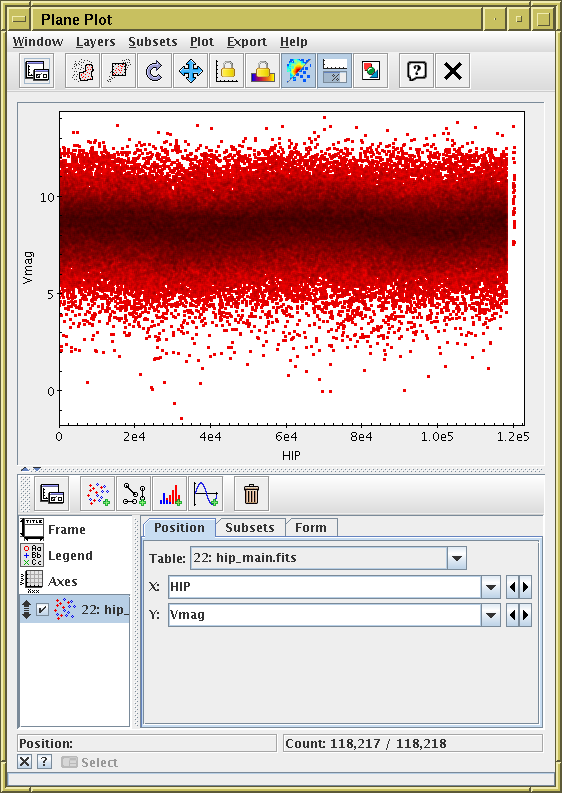}
\hspace*{0.5cm}
\includegraphics[width=0.506\textwidth]{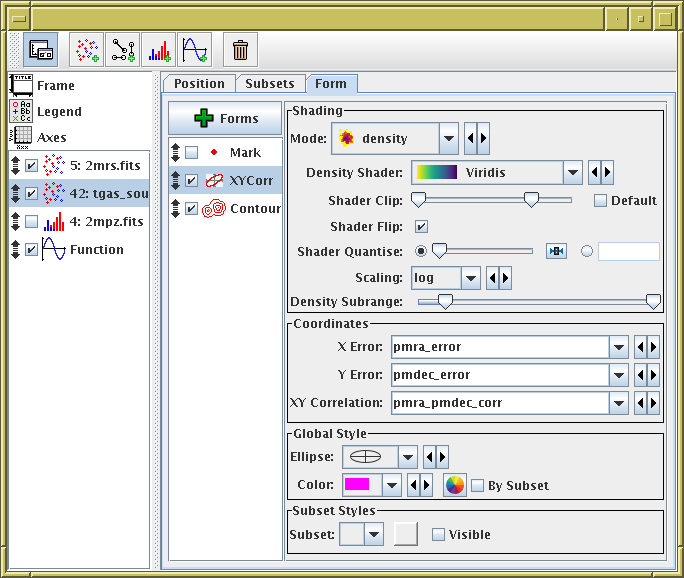}
\end{center}
\caption{
  TOPCAT's Plane Plot window.
  The left hand panel is what appears as soon as the user opens the window;
  some default plot is displayed.  It is easy to see how to change
  the data to be plotted.
  The right hand panel shows the control panel from the bottom of the
  plot window (here it has been expanded and floated out into its own window),
  as configured for controlling a much more complicated plot.
  Each of the controls towards the right can be adjusted interactively
  with instant effects on the displayed plot.
  The various tabs and list items provide more configuration options for
  other aspects of the plot.
}
\label{fig:plotwin}
\end{figure}

In practice, this means that if the user hits one of the top-level
{\em Plot\/} buttons in the application's main window,
a reasonable plot is shown, if at all possible.
If the user hits the {\em Plane Plot\/} button,
a~scatter plot is displayed of the two first
numeric columns of the currently-selected table;
an example is shown in Figure~\ref{fig:plotwin}.
The bounds of the plot region are automatically adjusted to
display all the points in the selected dataset.
The~shading mode ({\em Auto\/} from Figure~\ref{fig:plotmodes})
is one which works well for both small and large datasets.
This default plot is unlikely to be the one the user wants to see,
but it is easier to take an existing GUI and modify its default settings,
with instant visual feedback at every step,
than to be presented with a lot of blank fields to fill in
before any result is shown.
The controls visible in the same plot window make it obvious
how to choose a different table or different columns for the plot.
It is somewhat less obvious how to modify the plot by
changing marker characteristics, overplotting other datasets,
adding error bars or contours, changing axis scaling or annotation etc,
but by exploring the various tabs and list items the user
can explore and adjust the various options in as much detail as they require.
In this way each user can benefit from as much configuration effort
as they are willing to expend, rather than being scared off
by an initial need to understand the tool in detail.
Comprehensive documentation is provided for each feature,
accessible from a help button at the top of the window,
though it is more of a pleasant surprise than a general expectation
that confused users will seek and read this material.

This approach provides a GUI that is usable by novices to produce
basic plots, but which can also be exploited by experts for very
detailed control.
However, the question of providing a usable interface for configuring
complex plots is not a solved problem in TOPCAT;
it also becomes more acute as additional plot types and options are added.
While many users do manage to use the various combined features
to perform sophisticated visualisations it is probably the case that
only a minority understand the full range of available capabilities.

\section{Alternative Interfaces}
\label{sec:otherui}

This paper mainly discusses the GUI application TOPCAT.
However, it is possible to access the visualisation functionality
in a number of ways
from outside of that application, and many of the same remarks apply.

Alongside TOPCAT is a suite of command-line tools by the name of
STILTS (STIL Tool Set) \citep{2006ASPC..351..666T},
which provides scriptable access to
most of the functionality available from TOPCAT.
Both are based on the table access library
STIL (Starlink Tables Infrastructure Library),
and all of these items are developed and maintained,
with~some other partially related software, in a single group
of packages collectively known as Starjava.
These packages are available at the from the URLs
\url{http://www.starlink.ac.uk/stilts/} and
\url{http://www.starlink.ac.uk/stil/}.

STILTS provides commands that can generate all the visualisations
available from TOPCAT's GUI, and in fact most of the figures
in this paper were generated using STILTS,
since its scriptable nature makes it more suitable for
careful preparation of published figures
than TOPCAT's point and click operation.

The output of STILTS, like that of TOPCAT,
can be to various bitmapped or vector graphics file
formats (including PNG, GIF, PDF and PostScript)
or to an interactive window on screen that allows the same
mouse-controlled navigation actions as TOPCAT.

The classes used for visualisation in both cases
form a library known informally by the name {\em plot2}.
These classes have not so far been formally
packaged as a separate product,
but are contained within the STILTS jar file and can be used independently
of either the TOPCAT or STILTS applications,
to provide high performance static or interactive visualisation
within third party Java applications.
Depending on what functionality is required,
the code required for this is licensed under the LGPL or GPL.
Some more background on this possibility is described in
\cite{2015ASPC..495..177T}.

\begin{figure}[H]
\begin{center}
\includegraphics[width=0.7\textwidth]{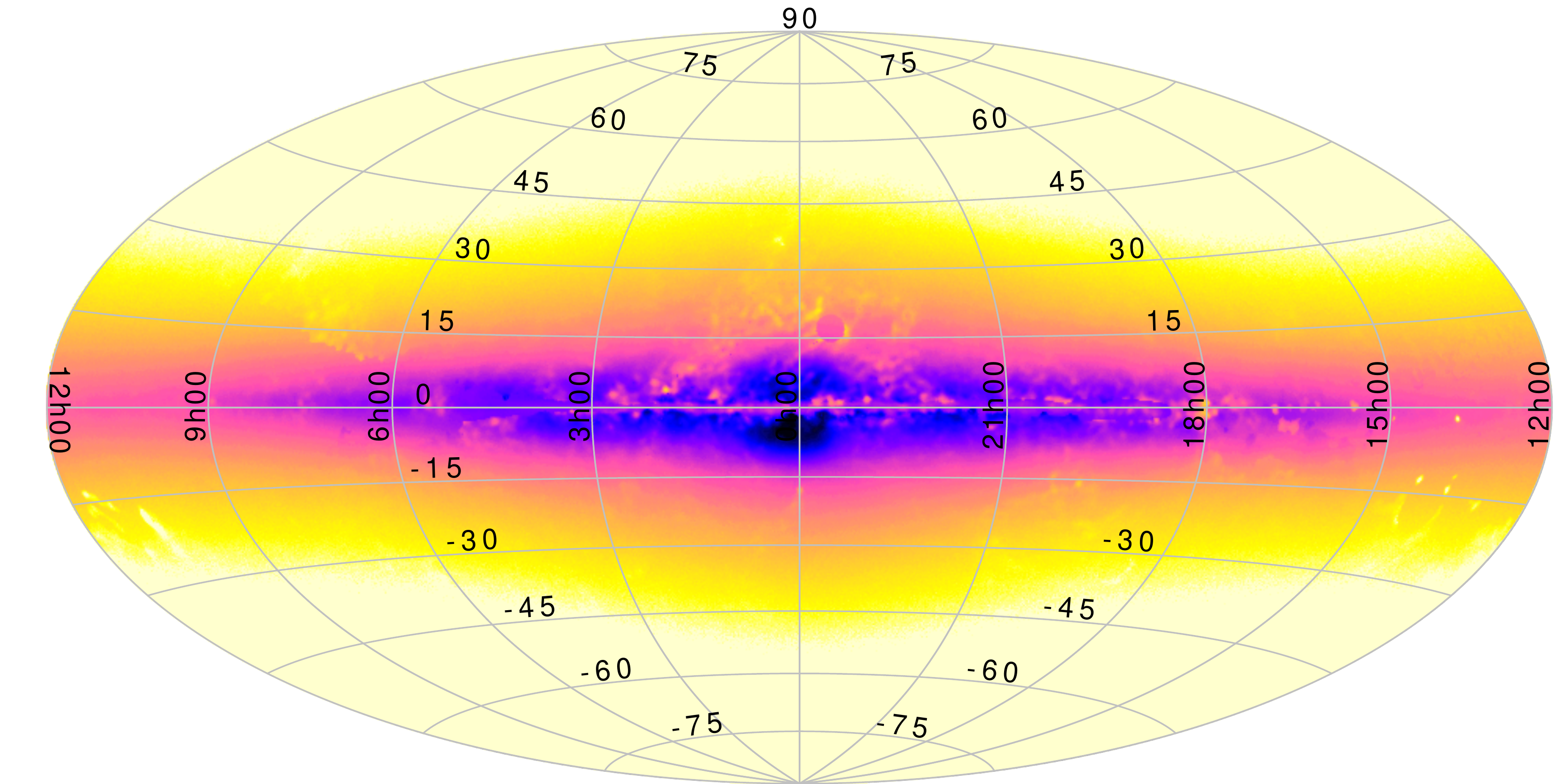}
\end{center}
\caption{Plot of a large table:
   A map representing simulated density on the sky of stars in the
   Milky Way \citep{2012A&A...543A.100R}.
   This figure (originally from \cite{2015ASPC..495..177T})
   was generated from a 2 billion row table in about 30 min
   on a normal desktop computer.
}
\label{fig:mw}
\end{figure}

As explained in Section~\ref{sec:scalability},
plotting from STILTS is actually more scalable than that from TOPCAT.
While there is no fixed limit on the size of tables loaded into TOPCAT,
it is not really intended for use with tables more than a few tens of
millions of rows; interactive use is generally sluggish on such data,
and in some cases memory usage can be high.
STILTS visualisation on the other hand is in most cases able to
stream data within a fixed (and quite small) memory footprint,
so it is possible to generate static plots of arbitrarily large
data sets quite easily.
Figure~\ref{fig:mw} shows an all-sky density plot generated
from a 2 billion row table in about 30 min.

\section{Implementation Notes}
\label{sec:performance}

The user experience-driven requirements of
usability with large datasets, fast navigation,
flexible configurability and instant visual feedback
place considerable constraints on the implementation.
In~this section we outline some of the strategies
in place to deliver these features.

\subsection{Scalability}
\label{sec:scalability}

The first requirement, built into the whole of the TOPCAT application
and its underlying libraries, is to be able to process
tables with large row counts $N_{\rm row}$
in fixed memory if at all possible.
This means that generating a plot must not,
unlike many off-the-shelf Java plotting libraries,
allocate an object or other storage for each input row or plotted point.
Instead, where possible, the plotting system uses data structures
that scale with the number of pixels rather than the number of rows.
This rule is violated in some cases, for instance if the number of
rows can be determined to be small, or in a few cases where it is unavoidable
such as $z$-stacking points for 3-d plots,
but most of the common plot types obey it.

To support this model, the various plot layer implementations
work with an abstraction of the data that simply iterates over
each row, returning typed values (typically double precision scalars,
though in some cases boolean, string or array values)
for the required coordinates at each step.
For~each iteration they can then either paint
directly to the graphics system
or populate some limited-size data structure
that will be used for graphics operations later in the rendering process.

The harnessing code can then decide how to deliver the iteration over
the data values from the original table.
The TOPCAT application reads the relevant values into in-memory
primitive arrays once it is known what coordinates are required,
ensuring maximal subsequent access speed, since~actually extracting these
values from the underlying loaded tables may be somewhat time-consuming.
This~does entail some $N_{\rm row}$-scale memory usage,
though usually at an acceptable level, e.g.,\ only 16 bytes per row
for a 2-d scatter plot.
However the STILTS application writing to a static image file
simply iterates over
the rows of the underlying table without intermediate caching,
thus requiring little additional storage.
These different strategies have different benefits:
TOPCAT, having prepared the data for a given visualisation
at the expense of some memory usage
can plot subsequent frames using the same data
(e.g.,\ the results of user navigation) as quickly as possible.
STILTS (in its default configuration) may take longer for each frame
of an animation sequence but can process arbitrarily large tables in a
small memory footprint.
In the common case where STILTS is painting to a single static frame,
the~benefits for subsequent replots would not be useful.

\subsection{Responsive User Interface}

As discussed in Section~\ref{sec:configui},
the visualisation user interface contains many controls,
each of which may change the appearance of some or all
parts of the plot; the axes or one or more of the data layers
contributing to a particular visualisation.
A responsive user interface requires that whenever one of these
controls is adjusted, the display is updated accordingly.
However, regenerating the whole plot from scratch may be expensive
for a large or complex plot, so this should be avoided where possible.
In some cases (e.g.,\ changing the plot colour of a currently hidden dataset)
perhaps no replot is required at all.
In other cases (e.g.,\ axis annotations only are changed)
some parts of the plot must be redrawn but the results of
previous computations could be re-used.
Or perhaps (e.g.,\ coordinate data is replaced by a different table column)
the whole thing needs to be redrawn.
In general various parts of the plotting computations can be cached,
and a great deal of effort goes into working out,
whenever the controls are adjusted, which computations from the
previous plot can be re-used.

The way this works is that every time a user control is adjusted,
it triggers a {\em replot\/} action.  This~calculates a set of {\em label\/}
objects for each of several plot characteristics such as the currently
visible region of parameter space,
the set of table data required for each plot layer,
the per-layer configuration style options etc.
Together, this set of labels completely characterises the plot.
These label objects are small and cheap to produce,
so multiple replot actions per second
(for instance, as a user drags a slider) are in themselves easy to service,
and this step can be done on Swing's Event Dispatch Thread (EDT)
without impairing the responsiveness of the overall application GUI.
If this is the first plot to be produced in a given window,
these labels are stored for later reference, and then
fed back to the plot components that generated them,
providing instructions to draw the plot which is then
calculated and displayed.
However, on subsequent plots in the same window,
the plotting system performs various
comparisons
of the labels for the new frame with those that specified the previous frame.
Specifically, Java's {\tt Object.equals}
   method is used for label comparison,
   so these label objects must be written with carefully implemented
   equality semantics.
If the set is exactly equivalent to that for the previous frame,
no replot needs to be done.
In general, some recalculation or redrawing will be needed,
but less than would be required for regenerating the whole plot from scratch.
This computation prepares a new plot bitmap on a worker thread,
which it passes on completion back to the EDT for display
in the plot window.
A queue of replot requests is maintained, and if a new one comes in
while another is waiting to be performed, the older one is discarded.
Whether requests in progress are aborted depends on some
logic to decide whether the new request looks like a minor (navigation)
or major (plot new data) configuration update.

The details of the selective caching that underlies this are quite
complicated, but an example may be illustrative.
A plot layer performs its plotting in two stages:
in the {\em planning\/} stage it is given the opportunity
to produce a {\em plan\/} object that may represent the results of
expensive computations,
and in the {\em painting\/} stage
it is given back the same plan to use in order to perform
actual graphical output.
Management, including optional caching, of the plan objects
is done by the plotting application and not the layer itself.
If the layer can determine that a plan
equivalent to the one it needs to produce
for the currently requested plot
is already available, because the management level has cached it
from an earlier invocation,
it can skip the planning stage and
use the previously calculated plan for the painting stage instead.
Hence: a density map layer might generate a plan containing a grid of
bins populated by the expensive work of iterating over the table rows,
and then in the painting stage simply transfer this grid to a
bitmap using some configuration-determined colour map.
If a subsequent invocation uses the same grid data
but a different colour map,
because the user has adjusted the colour map controls but not moved the grid,
it can repaint the image (cheap)
without requiring a rescan of the table data (expensive).
The result is that the user can
adjust colour map parameters with instant visual feedback
even for a large dataset.

\subsection{Configuration Option Management}

As discussed above, many configuration options are available
to control the data layers that combine to form a given visualisation,
along with the details of the axis representation and annotation,
legend display, plot dimensions, font selection etc.

In order to reduce the implementation complexity associated with
these hundreds of options, each one is represented by a standard
object known internally as a {\em ConfigKey}.
Each of these keys can supply user-directed metadata (name, description),
value type (which may just be a number or some more complex type like
a colour map or marker shape),
a sensible default value,
a GUI component for specifying values,
and methods for mapping between typed values and string representations.
It is important that the default values of all keys taken together
to specify a given plot will combine to give some reasonable
default plot as discussed in Section~\ref{sec:configui}.

Different components of the plotting system make use of these keys to
build the user interface and gather configuration information
without hard-coded knowledge of each plot and layer type.
A~harnessing application needs to establish which plot type and layers
are in use, interrogate them for their ConfigKeys, and 
then acquire values for each key that can
be fed back to the plot components to generate the plot.
In TOPCAT's case it sets up plot window controls by stacking the
relevant GUI components ready for adjustment by the user,
while STILTS interrogates the list of name-value pairs
supplied on the command line.
Other front ends based on name-value pairs have also been implemented,
including a cgi-bin interface for HTTP operation and
a Jython front-end to STILTS;
both are available and documented within the STILTS distribution
itself, as the STILTS {\tt server} task and the
JyStilts application respectively.

The user documentation for each plot type can also be generated
programmatically at documentation build time by interrogating
each key for its user metadata;
around 130 of the 400 pages in the PDF version of
the STILTS user document are
auto-generated from ConfigKey objects in this way.

The result of organising the configuration options in this uniform way
is that new configuration options can be introduced easily
by making only localised changes to plot type or layer type code;
no~corresponding updates to the UI code or hand-written
documentation are required.

\section{Use Beyond Astronomy}
\label{sec:nonastro}

TOPCAT has been developed for astronomers, with the support of
funding agencies whose responsibility is to the astronomy community.
The large majority of its use to date has been within astronomy,
mostly for use with source catalogues.
It is also applied to other types of astronomical table
such as time series and event lists,
and within some related but distinct fields such as
planetary and solar system science.

However, though it has much functionality that is specific to astronomy
(understanding of sky coordinate systems,
data access using astronomy file formats and
Virtual Observatory protocols,
table join techniques appropriate for the celestial sphere)
many of its capabilities are suitable for any kind of tabular data,
and some adventurous groups in other disciplines are also
making enthusiastic use of it.

\begin{figure}[H]
\centering
\includegraphics[width=0.75\textwidth]{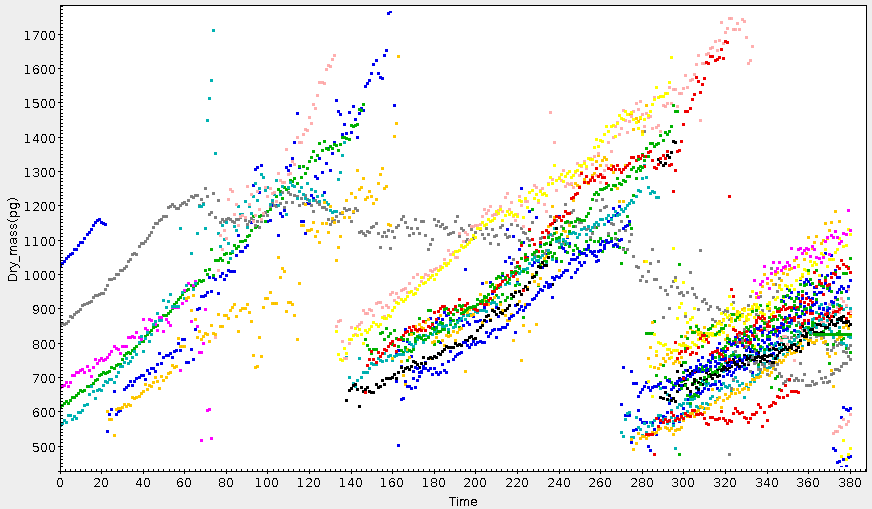}

\caption{
  Growth and division of cells over time.
  Cell mass, represented by one colour for each distinct cell,
  increases until mitosis, when it is replaced by two daughter cells.
  In this plot the grey cell has become ``blocked'',
  losing mass slowly instead of dividing.
  {\em Credit:\/} P.~Pognonec, Universit\'{e} Nice.
}
\label{fig:pognonec}
\end{figure}

\begin{figure}[H]
\begin{center}
\includegraphics[width=0.8\textwidth]{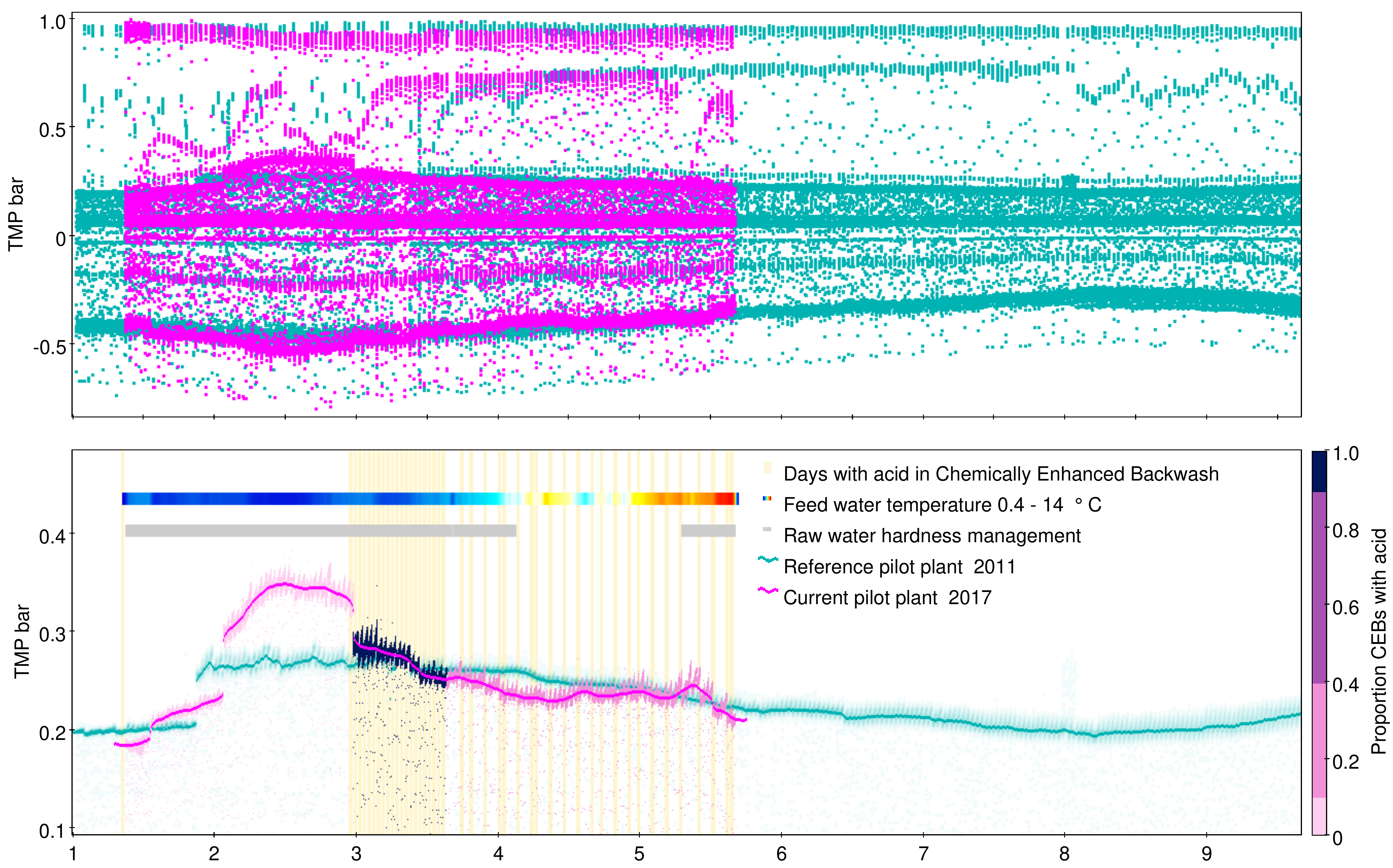}
\end{center}
\caption{
  Time series (decimal month January--September)
  of Trans Membrane Pressure over two different ultrafiltration
  pilot plants in Gothenburg Sweden (7.7M rows).
  { (\textbf{Top})\/}
  Two series of correct but noisy raw data,
  obviously not suitable for mean aggregations;
  { (\textbf{Bottom})\/}
  Exactly the same data as top display,
  clarified by use of subsets, transparency,
  quantile smoothers, marking by auxiliary \emph{Y} axis, histogram
  by time as event marking, and densograms for additional
  quantitative variable as well as operational information.
  {\em Credit:\/} H.~Rydberg, Sustainable Waste and Water, City of Gothenburg.
}
\label{fig:rydberg}
\end{figure}

One example is the group of P.~Pognonec from Universit\'{e} Nice,
who use it for work investigating
cellular events such as proliferation, mitosis and cell death.
They report TOPCAT as their preferred option for
visualising with dotplots and histograms the large tables
(ten million cells with 50--100~parameters)
of data produced by image analysis of high throughput microscopy
acquisitions.  An~example is shown in Figure~\ref{fig:pognonec}.
Another is operational and laboratory work by H.~Rydberg
in the company Sustainable Waste and Water, City of Gothenburg in Sweden,
where it is used especially for interactive analysis of long time
series data concerning water quality; see Figure~\ref{fig:rydberg}.
In this case the capability to ingest large raw datasets
without prior aggregation and navigate interactively
makes it possible to clean data
and identify trends over a wide range of timescales.
The author has also had other informal reports of TOPCAT's
sporadic use in
bioinformatics, finance, urban transport planning and
flight testing.

While the current funding arrangements do not prioritise support
outside of astronomy, the author is very interested to hear of
potential or actual uses in other domains,
and willing to supply modest support,
for either casual use or
adapting the application or underlying libraries
to other requirements.
One missing feature that should be noted when considering
applying the software more widely is its weak support for categorical data,
which is not very common in TOPCAT's core use cases.
However, enhancements in this area are possible in the future.

\section{Software Availability}
\label{sec:availability}

TOPCAT is written in pure Java, and distributed as a single jar file
depending only on the Java Standard Edition (Java SE), currently
version 6 or later.
The wide availability and excellent portability and backward compatibility
characteristics of the Java platform mean that it can therefore be installed
and run very easily on all widely used desktop and laptop computers.
The jar file, as well as a MacOS DMG file, can be downloaded
from the project web site \url{http://www.starlink.ac.uk/topcat/}.
Other information including comprehensive tutorial and reference
documentation,
full version history, pointers to mailing lists etc can be
found in the same place.
The package has also recently been made available as part of
the Debian Astro suite \citep{2016arXiv161107203S}.

The software is available free of charge under the GNU Public Licence,
and the source code is currently hosted on
github (\url{https://github.com/Starlink/starjava/}).

\section{Conclusions}
\label{sec:conclusions}

TOPCAT is a GUI application for manipulating tables,
that amongst other capabilities provides sophisticated
visualisation capabilities for tabular data.
It is a traditional desktop application, requiring neither
exotic hardware nor server support.
It has been developed within the context of astronomy
and is widely used in that field, but is suitable,
along with its command-line counterpart STILTS and
underlying Java libraries,
for visualising many other kinds of tabular data.
The focus is on highly configurable interactive plots of
both small and large (multi-million-row) tables,
offering many variations on the representation of point clouds
in one, two or three dimensions,
with the aim of revealing expected and unexpected relationships
at multiple scales in large and high-dimensional~datasets.

 \newpage
%%%%%%%%%%%%%%%%%%%%%%%%%%%%%%%%%%%%%%%%%%
\acknowledgments{Development of TOPCAT's current visualisation
capabilities has been supported by a number of grants from
the UK's Science and Technology Facilities Council.
The features described here have benefitted greatly from
advice, comments and feedback from its active user community.
Special thanks to Henrik Rydberg
and Philippe Pognonec for their input
on use in non-astronomical contexts.
The author also thanks the anonymous referees whose constructive
comments have improved the paper.
}

%%%%%%%%%%%%%%%%%%%%%%%%%%%%%%%%%%%%%%%%%%
\conflictsofinterest{The author declares no conflict of interest.
The funding sponsors had no role in the design of the study;
% is this really applicable/necessary here?
in the collection, analyses, or interpretation of data;
in the writing of the manuscript, and in the decision to publish the results.
} 

\externalbibliography{yes}
\bibliography{topcatviz}

%%%%%%%%%%%%%%%%%%%%%%%%%%%%%%%%%%%%%%%%%%
\end{document}